\pdfoutput=1
\documentclass[namedreferences]{SolarPhysics}
\usepackage[usenames]{color}
\usepackage[optionalrh,solaenum]{spr-sola-addons} 
\usepackage[T1]{fontenc}
\usepackage[latin1]{inputenc}
\usepackage{tikz}

\usetikzlibrary{arrows,snakes,backgrounds,patterns}
\usepackage{pgf}
\usepackage{url}             

\graphicspath{{figures//}}

\usepackage{savesym}
\savesymbol{iint}
\savesymbol{iiint}

\input{alphabet}
\newsavebox{\fminibox}
\newlength{\fminilength}

  \def\+{^\dagger}
\def\I{\,|\,}           

\def\nequiv{\not\kern-.05em\equiv}
\def\egal{\kern-.5em=\kern-.5em}        
\def\propt{\kern-.2em\propto\kern-.2em} 

\def\argmin{\mathop{\mathrm{arg\,min}}} 
\def\Argmin#1#2{\displaystyle \argmin_{#1}\left\{{#2}\right\}}  %

\def\intdouble{\int\kern-0.3em\int}
\def\inttriple{\int\kern-0.3em\int\kern-0.3em\int}

\def\rond#1{\overset{\kern-0.33em~_\circ}{#1}}
\def\rondit[#1]#2{\overset{\kern#1~_\circ}{#2}}

\newcommand{\degree}{\ensuremath{^\circ}}

\runningtitle{\textsf{TomograPy}: Solar Tomography}

\runningauthor{N. Barbey \textit{et al.}}

\begin{document}

\begin{article}

\begin{opening}

\title{\textsf{TomograPy}: A Fast, Instrument-Independent, Solar Tomography Software.}

\author{N.~\surname{Barbey}$^{1}$\sep
        C.~\surname{Guennou}$^{2}$\sep
        F.~\surname{Auch\`ere}$^{2}$
       }

       \institute{$^{1}$ SAp/Irfu/DSM/CEA, Centre d'\'etudes de
         Saclay, Orme des Merisiers, B\^atiment 709, 91191 Gif sur
         Yvette, France
         email: \url{nicolas.barbey@cea.fr}\\
         $^{2}$ Institut d'Astrophysique Spatiale, B\^atiment 121,
         Universit\'e Paris-Sud, 91405 Orsay, France
         email: \url{frederic.auchere@ias.u-psud.fr}
         email: \url{chloe.guennou@ias.u-psud.fr}\\
       }



\begin{abstract}
  Solar tomography has progressed rapidly in recent years thanks to the
  development of robust algorithms and the availability of more
  powerful computers. It can today provide crucial insights in solving
  issues related to the line-of-sight integration present in the data
  of solar imagers and coronagraphs. However, there remain 
  challenges such as the increase of the available volume of data, the
  handling of the temporal evolution of the observed structures, and
  the heterogeneity of the data in multi-spacecraft studies.
  We present a generic software package that can perform fast
  tomographic inversions that scales linearly with the number of
  measurements, linearly with the length of the reconstruction cube
  (and not the number of voxels) and linearly with the number of cores
  and can use data from different sources and with a variety of
  physical models:
  \textsf{TomograPy} (\url{http://nbarbey.github.com/TomograPy/}), an
  open-source software freely available on the Python Package
  Index. For performance, \textsf{TomograPy} uses a
  parallelized-projection algorithm. It relies on the World Coordinate
  System standard to manage various data sources. A variety of
  inversion algorithms are provided to perform the tomographic-map
  estimation. A test suite is provided along with the code to ensure
  software quality. Since it makes use of the Siddon algorithm it is
  restricted to rectangular parallelepiped voxels but the spherical
  geometry of the corona can be handled through proper use of priors.
  We describe the main features of the code and show
  three practical examples of multi-spacecraft tomographic inversions
  using STEREO/EUVI and STEREO/COR1 data. Static and smoothly varying
  temporal evolution models are presented.
  
\end{abstract}
\end{opening}



\section{Introduction}
\label{sec:introduction}

\subsection{Motivation}
Except for the rare case when {\it in-situ} exploration is
practicable, the properties of astronomical objects are deduced from
the analysis of the properties of light only. Most astrophysical
measurements are therefore affected by the problem of line-of-sight
(LOS) integration, \textit{i.e.} the modification of the signal of
interest by background and foreground emission and absorption. This
problem is one of the major sources of uncertainties in the
diagnostics of the solar plasma.


Integration along the LOS tends to confuse structures to the point
that measurements crucial to the understanding of coronal physics
are difficult to interpret. The controversy about the nature of polar plumes is
one example. Polar plumes are observed at visible and UV wavelengths
extending quasi-radially over the solar poles. Their appearance in
photographic records led to the classical view of plumes as being
pseudo-cylindrical structures denser than the surrounding
corona. However such linear features in the images can also result
from chance alignments of fainter structures distributed along a
network pattern and integrated along the LOS. Both types of plumes
have been supported by different authors, and it is possible that the
two types coexist. See, \textit{e.g.},~\inlinecite{gabriel2009} for a detailed
discussion. Since the two proposed types of plumes have nearly
identical properties in remote-sensing data, the true nature of these
objects remains a subject of controversy.

One can also cite the problem of background estimation in the coronal-loop
debate. As building blocks of the solar corona, loops have been
extensively studied. However, we still cannot answer fundamental
questions such as what processes are responsible for their formation
or for their heating. One of the factors explaining this state of facts is that
the determination of physical parameters such as
density and temperature within the loops is rendered difficult by LOS
superimposition. For example,~\inlinecite{terzo2010} have shown that
different estimations of the loop-background radiation lead to
different temperature profiles and different conclusions regarding the
loop cooling.

Different strategies have been devised over the years to overcome the
limitations imposed upon remote-sensing data by LOS integration. One
obvious approach is to select an observation time when the solar
corona presents a simple geometry for which it is possible to estimate
the contribution of the various regions of the LOS. When observing
polar plumes, for example, data are acquired preferentially at solar
minimum when the polar holes are well developed so that the contribution
of streamers to the foreground and background is minimum.  However,
even if these conditions are met, it is likely that several plumes or
plumes and inter-plumes are superimposed along the LOS, thus confusing
the interpretation. Therefore, favorable observing conditions are
generally not sufficient to exclude possible LOS ambiguity.

If a simple coronal configuration cannot be assumed, or more generally
if superimpositions cannot be ruled out, one has to devise means of
analyzing the LOS content. Spectroscopic techniques such as the
Differential Emission Measure (DEM) can be used to estimate the quantity
of emitting plasma along the LOS as a function of temperature. If this
approach is able to detect the presence of regions of different
temperatures along the LOS, it does not say how the temperatures are
distributed spatially. A single multithermal volume can have the same
DEM signature as the superimposition of several large-scale
isothermal structures.

Line-of-sight ambiguities can be alleviated, at least partially, if
one can make several simultaneous observations from different
locations. The twin spacecraft of the STEREO mission
\cite{kaiser2008stereo} were designed to achieve this. The two vantage
points that they offer provide precious information on the LOS
content. In some cases, especially with a high-contrast object having well-defined
boundaries such as coronal loops, direct stereoscopic
reconstructions can be performed. Such reconstructions can, for example,
be used to assess the quality of the background estimation used in
loop studies (\textit{e.g.}~\opencite{aschwanden2008}). However for more
diffuse objects not presenting sharp boundaries, or for which the
visible boundaries can be LOS-integration artifacts, such as
streamers, plumes, \textit{etc.}, direct, stereoscopic reconstruction is not
reliable. However, if two viewpoints or more are available, tomography
is a possible approach to inverting the LOS integration.

\subsection{Solar Tomography}
The term tomography encompasses a wide range of techniques aimed at
imaging the internal structure of objects. Tomographic techniques are
used in many areas of scientific research such as medicine,
geophysics, materials science, and astrophysics. In the particular case
of solar tomography, images recording the line-of-sight integration of
coronal emission and taken from different viewpoints are used to
estimate local physical quantities such as the electron number density
or temperature. This is achieved using computed tomographic
reconstruction techniques identical to the ones used in medical
computer tomography. Mathematically, it is an inversion of the
line-of-sight integration. This method is sometimes called Solar Rotational
Tomography (SRT) as it generally relies upon the solar rotation to
simulate data acquisition from different viewpoints.

However, there is significant differences between SRT and medical
computer tomography that renders the problem more difficult to solve
in the case of SRT. First, medical imaging scenarios benefit from high
signal to noise ratio (SNR) measurements and have much higher
measurement density than what is currently available with solar
observatories. But more importantly, medical imaging has much less
restriction on the number of point-of-view. Indeed, most of the time,
SRT is restricted to one instantaneous point-of-view, the only
exception being the STEREO twin spacecrafts. Another important
difference with medical imaging is the presence of an opaque sphere in
the middle of the region of interest: the photosphere. A similar issue
is the presence of occulter in coronograph instruments, which
restricts spatial information available in the data. This renders the
problem much more ill-posed in SRT than in medical imaging tomography.

Progress in solar tomography comes from the availability of new data,
new physical models, new inversion algorithms, and more powerful
computers.  Solar tomography can be traced back to
\inlinecite{van1950electron} who presented a one-dimensional inversion of
white-light data by fitting an analytic corona model.

Another seminal paper of SRT is
\inlinecite{altschuler1972determining}. It introduced computer aided
numerical estimation of the three-dimensional electron density of the
corona using data from the K-coronameter of the High Altitude
Observatory. The method reduces to a least-square estimation of the
coefficients of Legendre polynomials.

Later, \inlinecite{davila1994solar} investigated, through simulations, the
possibility of performing full three-dimensional (3D) solar tomography of
the corona. That article assumed that more than one spacecraft would
take data (up to nine actually) and thus did not require data to be taken
at different times. Emission-map estimation
was performed using the algebraic reconstruction technique (ART) which
is an iterative gradient method used to invert the linear tomographic
model. Estimated maps where reduced to $20 \times 20$ grids.

With the availability of the \textit{Solar and Heliospheric
  Observatory} (SOHO) data came the first three-dimensional maps of
the corona both in white light (\textit{e.g.}
\opencite{frazin2002tomography}) with the \textit{Large Angle
  Spectrometric Coronagraph} (LASCO: \opencite{brueckner1995lasco})
and in the ultraviolet (\textit{e.g.} \opencite{panasyuk1999three})
using the \textit{UltraViolet Coronagraph Spectrometer} (UVCS:
\opencite{kohl1995}). An algorithm-oriented article by
\inlinecite{frazin2000tomography} introduced a penalized likelihood
approach minimized using an iterative solver (conjugate-gradient) to
allow noise mitigation through proper regularization and modeling of
the outliers.

Several generalizations were then developed. For example,
\inlinecite{wiegelmann2003magnetic} introduced a method for the joint
estimation of the electron density and the magnetic field while
\inlinecite{frazin2009} proposed a reconstruction of the local DEM
from EUV images.

The temporal evolution of coronal structures during data
acquisition is one of the main issues in coronal tomography. Two
different directions have been investigated to address this issue:
either assume slow evolution (\textit{e.g.} \opencite{butala2010dynamic}) or
further restricting the possible evolution to specific structures
(\textit{e.g.} \opencite{barbey2008time}).

\subsection{Outline}
In this article we describe \textsf{TomograPy}, an open-source software package
implementing the main desirable features in a generic solar
tomography code (see, \textit{e.g.}, \opencite{frazin2005rotational} for a
review): capability to use both EUV and white-light data to estimate
the local electron density and temperature, modeling the temporal
evolution of structures during data acquisition, and performing rotational
tomography with multiple spacecraft, \textit{i.e.} with STEREO data.

In Section~\ref{sec:tomographic_inversion} we formulate mathematically
the tomographic inversion problem and introduce the notations used in
the description of the code given in
Section~\ref{sec:features}. Section~\ref{sec:performances} gives an
overview of the numerical performance of the code while
Section~\ref{sec:exemples} describes three practical examples of
tomographic reconstructions.

\section{Tomographic Inversion}
\subsection{Linear Inverse Problem}
\label{sec:tomographic_inversion}
The problem to invert can be expressed by Equation (\ref{eq:linear})
where $\yb$ is the data, $\Ab$ is called the physical model, $\xb$ is
the object map to estimate, and $\nb$ is an additive noise (which we
will assume Gaussian, independent and identically-distributed
(iid))
\begin{equation}
  \yb = \Ab \xb + \nb
  \label{eq:linear}
\end{equation}

The physical model represents all of the transformations that link the
quantity to estimate (\textit{e.g.} the local emissivity, electron
temperature, electron density, \textit{etc.}) to the data. It always
includes the line-of-sight integration and the model of temporal
evolution, even if this later is implicitly static. The physical model
may also include the formation process of the observed lines if one
wants to estimate a physical quantity such as the local electron
density or temperature instead of the local emissivity. We restrict
ourselves to cases where the data are a linear function of the unknown
quantities to determine. It is worth noting, however, that tomographic
inversion can be done even if the physical model is non-linear,
although accompanied by an important increase in complexity. An
exemple of non-linear tomography application is electrical capacitance
tomography \inlinecite{soleimani2005nonlinear} which is intrinsically
non linear. Using Monte Carlo Markov Chain (MCMC) methods, it would be
feasible to fit non-linear models of Coronal Mass Ejections as the one
provided in \inlinecite{thernisien2009forward}.

\subsection{Bayesian View on Linear Inversion}
In the Bayesian paradigm, a probability density function (PDF) is
associated with each variable. To invert the class of problems
described by Equation (\ref{eq:linear}), one needs to know the
statistical properties of the noise [$\nb$], which gives the
likelihood. One also needs to define a prior on the unknowns [$\xb$]: it
gives the PDF of $\xb$ knowing the data [$\yb$], which is called the
posterior on $\xb$. This is done through Bayes' rule given in
Equation (\ref{eq:bayes})
\begin{equation}
  f(\xb | \yb, \mathcal{M}) = \frac{f(\yb | \xb, \mathcal{M})
    f(\xb | \mathcal{M})}{f(\yb | \mathcal{M})}
  \label{eq:bayes}
\end{equation}

\noindent
where $\mathcal{M}$ regroups all of the assumptions on the model. In this
article, the PDF of $\xb$ given $\yb$ is noted $f(\xb | \yb)$.

In the case of a Gaussian multivariate likelihood and prior, the
posterior is also Gaussian, and thus fully determined by its mean and
covariance matrix. This is summed up in Equation
(\ref{eq:gaussian_bayes})
\begin{eqnarray}
  \label{eq:gaussian_bayes}
  f(\xb | \mathcal{M}) &\sim& \mathcal{N}(\zerob, \sigma_\xb^{2} (\Bb^T\Bb)^{-1}) \nonumber \\
  f(\yb | \xb, \mathcal{M}) &\sim& \mathcal{N}(\Ab\xb, \sigma_\nb^{2} \Ib) \nonumber \\
  f(\xb | \yb, \mathcal{M}) &\sim& \mathcal{N}(\hat{\xb}, \hat{\Sigmab}) \nonumber \\
  \hat{\xb} &=& \hat{\Sigmab} \Ab^T \yb \\
  \hat{\Sigmab} &=& \left( \Ab^T \Ab + \lambda \Bb^T\Bb \right)^{-1} \nonumber \\
  \lambda &=& \frac{\sigma_\xb^2}{\sigma_\nb^2} \nonumber
\end{eqnarray}

\noindent
where $\mathcal{N}(\mub, \Sigmab)$ is a multivariate Gaussian of mean
$\mub$ and covariance $\Sigmab$. In Equation (\ref{eq:gaussian_bayes})
we assumed an independent Gaussian noise of variance $\sigma_n^2$, so
that the covariance of the likelihood is $\sigma_n^2 \Ib$ where $\Ib$
is the identity matrix. We defined a zero-mean prior [$f(\xb |
\mathcal{M})$] with $\Bb$ being a prior model. $\Bb$ can be, for
instance, a finite-difference operator. A zero-mean prior combined
with a finite-difference operator means that the finite difference of
the map tends to be close to zero. In other words, this prior would
favor smoother solutions over non-smooth solutions close to the
data. This is a sensible choice for electron density at the scales
considered. $\sigma_n$ and consequently $\lambda$ are free
parameters. It is possible to assign a PDF to $\lambda$ in order to
estimate this parameter in an unsupervised way but it results
generally in very resource-consuming algorithms. In this article, we
will restrict ourselves to a fixed $\lambda$ in a supervised way.

Characterizing the solution in a Bayesian way requires the estimation
of both $\hat{\Sigmab}$ and $\hat{\xb}$. $\hat{\xb}$ gives the most
probable solution and $\hat{\Sigmab}$ gives information about the
uncertainties in the unknowns. However, in most practical cases, the
covariance matrix [$\hat{\Sigmab}$] is too large to be stored in
memory, and one only keeps $\hat{\xb}$. In this case, a full matrix
inversion is not required, and one can estimate $\xb$ much faster
using iterative schemes such as the conjugate gradient method.

Since $\xb$ is the maximum \textit{a posteriori} (MAP) of the problem,
it is also the minimum of the co-log--likelihood as written in Equation
(\ref{eq:criterion})
\begin{eqnarray}
  \label{eq:criterion}
  \hat{\xb} &=& \Argmin{\xb}{J(\xb)} \nonumber \\
  &=& \Argmin{\xb}{- \log\left[ f(\xb | \yb, \mathcal{M}) \right]} \nonumber \\
  &=& \Argmin{\xb}{- \log\left[ f(\yb | \xb, \mathcal{M}) f(\xb | \mathcal(M)) \right]} \\
  &=& \Argmin{\xb}{\left\|\yb - \Ab\xb \right\|^2 + \lambda \left\| \Bb \xb \right\|^2} \nonumber \\
  &=& \left( \Ab^T \Ab + \lambda \Bb^T\Bb \right)^{-1} \Ab^T \yb \nonumber
\end{eqnarray}

The term $\| \yb - \Ab \xb \|^2$ is a simple least-squares term. It
defines the closeness to the data. The second term [$\|\Bb \xb\|$] is
a regularization term that prevents the estimate from being noisy. In
terms of matrix inversion, $\Ab^T\Ab$ is ill-conditioned and
$\Bb^T\Bb$ is added in order to have a better-conditioned matrix. To
find $\hat{\xb}$, iterative-gradient methods need only the definition
of the criterion $J(\xb)$ and its gradient $\nabla_\xb J$. Gradient
methods can be order of magnitudes faster than the full inversion of
the matrix, especially when A and B are sparse or when the problem has
been properly preconditioned.

\section{Main Features of \textsf{TomograPy}}
\label{sec:features}

\subsection{Fast Parallelized Projector}
\label{subsec:projector}
\textsf{TomograPy} is a \textsf{Python} \cite{van1995python} package
build around a \textsf{C} implementation of the Siddon algorithm
\cite{siddon85} of line-of-sight integration, and is thus restricted
to rectangular parallelepiped voxels. This C projector has been
parallelized using OpenMP \cite{dagum2002openmp}. \textsf{Numpy}
\cite{oliphant2006guide} is a requirement as well as \textsf{PyFITS}
\cite{barrett1999pyfits} to handle FITS data files. Optionally, one
can use \textsf{SciPy} \cite{scipy} sparse matrix optimization
routines to perform fast linear inversions.
The algorithm has been carefully optimized using meta-programming
techniques to avoid \textsf{if} statements and function pointers in
the inner loop. This has been done using templates of \textsf{C} code,
and replacing key values in the source template to generate variations
in the \textsf{C} sources for various application (e.g.: float and
double values, projection and backprojection, presence of an obstacle
or not). Here the word template is not to be confused with
\textsf{C++} templates but is more closy related to the notion of web
template. The same idea is used in Numpy itself and allow more
flexibility to pure \textsf{C} code.

The projection algorithm provided with \textsf{TomograPy} can be used
with a variety of estimation methods as long as they rely on the
linear-operator interface. It allows for fast testing of various
optimization strategies. Results presented in this article will
exclusively use conjugate gradient schemes, but \textsf{TomograPy}
provides other options. The key requirement is for the algorithm to
rely on matrix--vector operations.

See Section \ref{sec:performances} for an analysis of the performance
and scaling of this implementation.
The \textsf{TomograPy} projector is well tested and provided with a test
suite, which fully covers this part of the code.

\subsection{Instrument Independence}
\textsf{TomograPy} takes as input FITS files (Flexible Image Transport
System: \opencite{wells81}) containing fully calibrated images
expressed in units consistent with the physical model chosen by the
user. \textsf{TomograPy} internally uses the World Coordinate System
(WCS) \cite{calabretta2002representations,greisen2002representations}
standard keywords to determine the position of the observer and to
define the projector from the data and the desired format of the
object map. \textsf{TomograPy} will therefore accept any data compliant
with the WCS standard. As the data of most current instruments are
already provided as WCS-compliant FITS files, all that is required is
to store a set of calibrated files in a directory that \textsf{TomograPy}
will be pointed to. For data that does not conform to WCS, it is
straightforward to write a small wrapper that will handle the
instrument-specific metadata and convert them into the corresponding
WCS keywords.

\textsf{TomograPy} allows inversions using data from multiple spacecraft,
for example with STEREO-A and B and SOHO. The data from the different
instruments nonetheless need to be consistent, \textit{i.e.} to record
the same physical quantity.

\subsection{Physical models}
As stated in section \ref{sec:tomographic_inversion}, any linear model
can be inverted using the same framework. \textsf{TomograPy} provides
with the possibility to perform inversions with the several models
described in this section. The \textsf{TomograPy} projector already
discussed in Section~\ref{subsec:projector} is a building block for all
of the models described here. We will first describe models of
temporal evolution and then models of line emission. It is possible to
combine these models to perform, for example, multi-spacecraft,
smooth, temporal rotational tomography. In the future, it will be
possible to combine the models presented here with models not yet
available in \textsf{TomograPy} such as the Differential Emission Measure
model \cite{frazin2009} or magnetic-field models
\cite{wiegelmann2003magnetic}.

\subsubsection{Single Spacecraft Static Tomography}
\label{subsec:substatic}
This is the simplest case. In the next two sections we will consider
only the line-of-sight inversion without assumptions on the line
formation process. In this case, static rotational tomography of the
solar corona can easily be formulated as in Equation (\ref{eq:linear})
once discretized. In this article, we will assume that the object-map
cube has been discretized using contiguous rectangular parallelepiped
voxels of identical shape; it is a requirement of the Siddon
algorithm. Since the intensity on one detector results from the
integration along the line of sight of the emission in the observed
object, it can be expressed as in Equation (\ref{eq:lo}):
\begin{equation}
  y_j = \sum_i P_{i,j} e_i + n_j
  \label{eq:lo}
\end{equation}

\noindent
where $y_j$ is the intensity of the detector $j$, $e_i$ is the
emission in the voxel $j$, and $P_{i,j}$ is the length of the segment
of the line of sight $i$ that corresponds to the voxel $j$, and $n_j$
is the noise observed on detector $j$. Reformulating Equation
(\ref{eq:lo}) in terms of vectors and matrices, we obtain Equation
(\ref{eq:lo_t}) where the $t$ index refers to the time at which the
data have been taken.
\begin{equation}
  \yb_t = \Pb_t \eb_t + \nb_t
  \label{eq:lo_t}
\end{equation}

\noindent
where $\Pb$, the \textsf{TomograPy} projector, is the most basic block for
building physical models. Note that we indexed both axes of the
detector using one index as well as the voxels of the object map. We
can do the same on a time index, since we assume that there is no
temporal evolution ($\eb$ does not vary with $t$). Regrouping all of the
data at all considered instants [$t$], results in Equation
(\ref{eq:srt}) which is similar to Equation (\ref{eq:linear}).
\begin{equation}
  \yb = 
  \left( \begin{array}{c} \yb_1 \\ \vdots \\ \yb_T \end{array} \right) =
  \left( \begin{array}{c} \Pb_1 \\ \vdots \\ \Pb_T \end{array} \right) \eb
  + \left( \begin{array}{c} \nb_1 \\ \vdots \\ \nb_T \end{array} \right) =
  \Pb \eb + \nb
  \label{eq:srt}
\end{equation}

This model can be used to estimate emission maps from EUV data using
formula (\ref{eq:criterion}). We can use a smoothness prior to avoid
having too much noise in the maps. This is done using a
finite-difference operator along each axis of the maps for the $\Bb$
matrix.  If we want to account for the lower signal-to-noise ratio
that we typically have in solar rotational tomography, we can have
finite-difference operators weighted by the altitude of the considered
voxels. Finally, we have the following equation
\begin{equation}
  \label{eq:srt_inversion}
  \hat{\eb}_{SRT} = \Argmin{\eb}{\left\|\yb - \Pb\eb \right\|^2 + \lambda \left\| \Db \Rb \eb \right\|^2}
\end{equation}
\noindent
where $\Db$ is the finite-difference operator and $\Rb$ is a diagonal
operator with the height of the voxels on the diagonal. The use of a
smoothness prior increasing linearly with height, allows for the maps
not to be affected by the difference between spherical grids and
Cartesian grids. Spherical grids have bigger voxels at high altitudes
increasing the SNR per voxel with height. This is not the case for
Cartesian grids but it is compensated by the use of a height-dependent
prior.

\subsubsection{Multiple-Spacecraft Static Tomography}
\label{subsec:multistatic}
If now we want to use data from multiple spacecraft, with the static
assumption we can use Equation (\ref{eq:two_spacecraft}) (assuming two
spacecraft A and B without loss of generality).
\begin{eqnarray}
  \yb_A &=& g_A\Pb_A \eb + \nb_A \nonumber \\
  \yb_B &=& g_B\Pb_B \eb + \nb_B
  \label{eq:two_spacecraft}
\end{eqnarray}

Each of Equations (\ref{eq:two_spacecraft}) are derived from
(\ref{eq:srt}) but both spacecraft can have different gains ($g_A$ and
$g_B$) at the  wavelength considered. In this model, it is not possible
to assume a different behavior of the filters as a function of the
wavelength since $\eb$ needs to correspond to emission integrated in
one filter. Fortunately, this assumption is valid to a good
approximation for several existing instruments. The passbands of the
two \textit{Extreme UltraViolet Imagers} (EUVI: \opencite{wuelser2004euvi}) on
STEREO and those of the \textit{Extreme ultraviolet Imaging Telescope} (EIT:
\opencite{delaboudiniere1995eit}) have for example been designed to be
identical.

Equation (\ref{eq:two_spacecraft}) can be reformulated as Equation
(\ref{eq:linear}) by a simple concatenation as shown in Equation
(\ref{eq:lin2space}).
\begin{equation}
  \yb_{A,B} = 
  \left(\begin{array}{c} \yb_A \\ \yb_B \end{array}  \right) = 
  \left(\begin{array}{c} g_A\Pb_A \\ g_b\Pb_B \end{array}  \right) \eb + 
  \left(\begin{array}{c} \nb_A \\ \nb_B \end{array}  \right) = 
   \Pb_{A,B} \eb + \nb_{A,B}
  \label{eq:lin2space}
\end{equation}

Finally, multiple-spacecraft tomography in the static case can be
formulated as single-spacecraft tomography as long as all of the
instruments have the same spectral bandwidth. The only modification is
the multiplication by gains, which vary from one instrument to the
other. We can then write the estimated map using Equation
(\ref{eq:criterion}) as in Equation (\ref{eq:srt_multiple_inv}).
\begin{equation}
  \label{eq:srt_multiple_inv}
  \hat{\eb}_{SRT,A,B} = \Argmin{\eb}{\left\|\yb_{A,B} - \Pb_{A,B}\eb \right\|^2 + \lambda \left\| \Db \Rb \eb \right\|^2}
\end{equation}

Using a DEM model, it is possible to extend multiple-spacecraft
tomography to cases in which the spacecraft have different bandpasses
since the spectral response is then integrated into the model. This is
the approach followed by \cite{frazin2009}.

\subsubsection{Smooth temporal tomography}
\label{subsec:smooth}
Because the corona is not static, dynamic models are desirable. In
this case however, it is no longer possible to simplify Equation
(\ref{eq:lo_t}) as in Equation (\ref{eq:srt}). When temporal evolution
is present, the recording of data taken at different times can be
expressed as
\begin{equation}
  \yb = 
  \left(\begin{array}{c} \yb_1 \\ \vdots \\ \yb_T  \end{array} \right) =
  \left( \begin{array}{ccc}
    \Pb_1 &  & \zerob \\
     & \ddots & \\
    \zerob & & \Pb_T
  \end{array} \right)
  \left( \begin{array}{c} \eb_1 \\ \vdots \\ \eb_T \end{array} \right) + 
  \left( \begin{array}{c} \nb_1 \\ \vdots \\ \nb_T \end{array} \right) =
  \Pb_{\mathcal{T}} \eb_{\mathcal{T}} + \nb
  \label{eq:srt_t}
\end{equation}

This results in a highly underdetermined inverse problem since there
are $T$ times more unknowns than in Equation (\ref{eq:srt}). This
underdetermination can be mitigated using either priors, such as a
temporal smoothness prior, or a parameterization of the temporal
evolution. Thus a classic, smooth, temporal solar-rotational
tomography (STSRT) would perform conjugate gradient estimation using
the criterion given in Equation (\ref{eq:stsrt}), where $\Db_r$ and
$\Db_t$ are the finite-difference operators in space and
time. Typically, due to strong underdetermination, the hyperparameter
$\lambda_t$ would be greater than $\lambda_r$, favoring solutions with
small temporal changes.
\begin{equation}
  \hat{\eb}_{STSRT} = \left\| \yb - \Pb_{\mathcal{T}}\eb_{\mathcal{T}} \right\|^2 +
  \lambda_r \left\| \Db_r \eb_{\mathcal{T}} \right\|^2 +
  \lambda_t \left\| \Db_t \eb_{\mathcal{T}} \right\|^2
  \label{eq:stsrt}
\end{equation}

This kind of approach using spatio-temporal regularization has been
explored before \cite{zhang2005analytical,khalsa2007resolution}.

Following the same approach as in Section~\ref{subsec:multistatic}, it
is possible to generalize this expression to the case of multiple
spacecraft.

\subsection{EUV Lines}
In the case of EUV lines or EUV bands, the dominant formation process
of the observed radiation is excitation by collisions between ions and
electrons. The local emissivity can thus be supposed to be isotropic
in which case the quantity inverted in Sections \ref{subsec:substatic}
to \ref{subsec:smooth} is directly the local emissivity of the plasma
summed over the spectral response of the instrument. Resonant
scattering may contribute significantly to EUV bands such as the 17.1
nm and 19.5 nm bands used in, \textit{e.g.}, TRACE, EIT, EUVI, and
AIA~\cite{schrijver2000}. If this is confirmed, then the local
emissivity is not isotropic and one needs to apply a correction factor
to the inverted quantities to deduce plasma emissivities.

\subsection{White Light: Thomson Scattering}
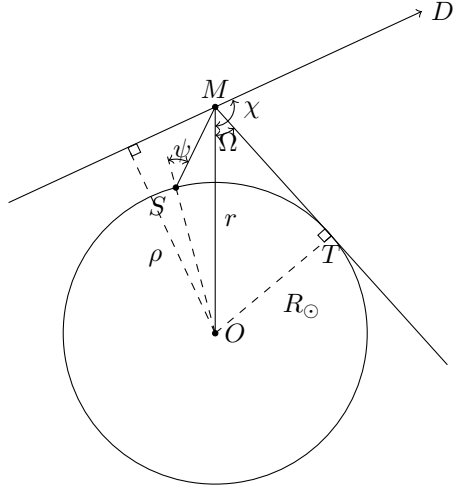
\begin{figure}[]
  \centering
  \begin{tikzpicture}[scale=.5]
  \coordinate (O) at (0,0);
  \draw (O) node[right] {$O$};
  \draw[fill] (O) circle (2pt);
  \coordinate (M) at (0,6);
  \draw (M) node[above] {$M$};
  \draw[fill] (M) circle (2pt);
  \coordinate (T) at (40.2:4);
  \draw (T) node[below] {$T$};
  \coordinate (S) at (105:4);
  \draw (S) node[below left,fill=white] {$S$};
  \draw[fill] (S) circle (2pt);

  \coordinate (S2) at (105:5);
  \coordinate (T2) at (barycentric cs:M=-1,T=2);
  \coordinate (D2) at (115:5.43);

  \draw (O) -- (M)
  node[right,midway] {$r$};
  \draw (O) circle (4);
  \draw[dashed] (O) -- (T)
  node[below right,midway] {$R_\odot$};
  \draw[dashed] (O) -- (S2);
  \draw (M) -- (S);
  \draw (M) -- (T2);
  \draw [->] (M) -- ++(25:6) node[right] {$D$};
  \draw [-] (M) -- ++ (205:6);
  \draw[dashed] (O) -- (D2)
  node[below left,midway] {$\rho$};

  \draw [<->]  (M) + (0,-.75) arc (-90:-48:.75);
  \draw (M) +(-70:1) node {$\Omega$};
  \draw [<->] (S) + (105:.75) arc (105:64:.75);
  \draw (S) +(80:1) node {$\psi$};
  \draw [<->] (M) + (-90:.5) arc (-90:22:.5);
  \draw (M) + (-10:.5) node [right] {$\chi$};

  \draw (T) -- ++ (132:.25) -- ++ (222:.25) -- ++ (313:.25) -- (T);
  \draw (D2) -- ++ (25:.25) -- ++ (-65:.25) -- ++ (-155:.25) -- (D2);

\end{tikzpicture}
  \caption
  {Geometry of Thomson scattering in the corona.  $M$ is the location
    of the scattering electron, $O$ is Sun center, $S$ is an
    emission point on the solar surface, $D$ is the observer, and $T$
    is the tangent to the solar surface. $\rho$ is the impact parameter
    of $(DM)$, $r$ is the distance of the scattering point to Sun
    center, $R_\odot$ is the solar radius. $\Omega$ is the angle between
    the line of sight and the tangent to the solar surface passing
    through the scattering point.}
  \label{fig:thomson_corona}
\end{figure}

In the case of white-light detectors, the measured intensity is
largely dominated by Thomson scattering of the photospheric radiation
by free coronal electrons. Figure \ref{fig:thomson_corona} shows the
geometry of Thomson scattering in the corona.  Following
\inlinecite{billings66}, the equations of Thomson scattering in the
corona are

\begin{equation}
  \begin{array}{rcl}
   I_t &=& 
   \frac{\pi\sigma}{2} \frac{I_0}{r^2} n_e \left[ (1 - u)C_3 + uC_4 \right] \\
   I_t - I_r &=& 
   \frac{\pi\sigma}{2} \frac{I_0}{r^2} n_e \frac{\rho^2}{r^2}
    \left[ (1 - u)C_1 + uC_2 \right] \\
  \end{array}
  \label{eq:thomson_corona}
\end{equation}

\begin{equation}
  \begin{array}{rcl}
    C_1 &=& \cos \Omega \sin^2 \Omega \\
    C_2 &=& -\frac{1}{8} \left[ 1 - 3\sin^2\Omega -
      \frac{\cos^2 \Omega}{\sin\Omega}(1 + 3 \sin^2 \Omega) 
      \ln \left( \frac{1 + \sin\Omega}{\cos\Omega} \right) \right] \\
    C_3 &=& \frac{4}{3} - \cos\Omega - \frac{\cos^3\Omega}{3}\\
    C_4 &=& -\frac{1}{8} \left[ 5 + \sin^2\Omega -
      \frac{\cos^2 \Omega}{\sin\Omega}(5 -  \sin^2 \Omega) 
      \ln \left( \frac{1 + \sin\Omega}{\cos\Omega} \right) \right]
  \end{array}
  \label{eq:4C}
\end{equation}

\noindent
where $u$ accounts for the center-to-limb variation and is a function
of wavelength, $I_t$ and $I_r$ are intensities in the radial and
transverse directions, $n_e$ is the electron density, $r$ is the
distance of the scattering point to the center of the Sun, $\rho$ is
the impact parameter of the line of sight, $\Omega$ is the angle
between the line of sight and the tangent to the solar surface passing
through the scattering point, $\sigma$ is the Thomson-scattering
cross-section and $I_0$ is the incident intensity.

The important point in these equations is that the intensity is a
linear function of the electron density. It is thus possible to
directly estimate the electron density using solar tomography on
white-light data. Note also that the Thomson-scattering equations can
be separated into coefficients that depend on the position [$M$] in
the corona (through $r$ and $\Omega$) and the line of sight (through
the impact parameter [$\rho$]). These coefficients are given in
Equation (\ref{eq:thomson_lo}).
\begin{equation}
  \begin{array}{ccc}
    o_k &=& \rho^2_k \\
    m_j &=& \frac{(1-u)C_1(r_j) + C_2(r_j)}{r_j^4} \\
    m'_j &=& \frac{(1-u)C_3(r_j) + C_4(r_j)}{r_j^2} \\
  \end{array}
  \label{eq:thomson_lo}
\end{equation}

Measurements are generally decomposed into a polarized brightness [pB]
component and a total brightness [B] component. Equation
(\ref{eq:brightness}) gives the equations for those quantities.
\begin{eqnarray}
  pB_k &=& \frac{\pi\sigma}{2}I_0 \frac{\rho_k^2}{r_j^4}
  \left[(1-u)C_1(r_j)+uC_2(r_j) \right] n_{e,j} \nonumber \\
  B_k &=& \frac{\pi\sigma}{2}I_0 \left\{
    - \frac{\rho_k^2}{r_j^4} \left[(1-u)C_1(r_j)+uC_2(r_j) \right] \right. \nonumber \\
    && \left. + 2\frac{1}{r_j^2} \left[(1-u)C_3(r_j)+uC_4(r_j) \right]
  \right\} n_{e,j}
  \label{eq:brightness}
\end{eqnarray}

From Equations (\ref{eq:thomson_lo}) and (\ref{eq:brightness}) one can
build linear direct models for pB and B images as in Equation
(\ref{eq:thomson_direct}), where $\nb_e$ is the discretized
electron-density map and $\ib_{pB}$ and $\ib_B$ are the pB and B
images respectively. $\Ob$, $\Mb$, and $\Mb'$ are diagonal matrices
with $o_k$, $m_j$, and $m'_j$ on their diagonals.
\begin{eqnarray}
  \ib_{pB} &=& \Tb_{pB} \nb_e + \nb \nonumber \\
  &=& \Ob \Pb \Mb \nb_e + \nb \nonumber\\
  \ib_B &=& \Tb_{B} \nb_e + \nb \nonumber \\
  &=& \left( \begin{array}{cc} -\Ob &\Ib \end{array} \right)
  \left( \begin{array}{cc} \Pb &\zerob \\ \zerob & \Pb \end{array} \right)
  \left( \begin{array}{c} \Mb \\ \Mb' \end{array} \right) \nb_e + \nb
  \label{eq:thomson_direct}
\end{eqnarray}

Note that almost twice as much computation is required for simulation
or inversion of total brightness data than for polarized brightness
data. A recent application of this model can be found in
\inlinecite{frazin2010three}.

\section{Performance}
\label{sec:performances}

We performed tests on a set of 64 images of 256 $\times$ 256 pixels,
and a reconstruction cube of 128 $\times$ 128 $\times$ 128 voxels. We
always use these parameters unless specified otherwise. For these
tests there is no obstacle, meaning that the ray tracing is not
interrupted as it would have been with the use of a model with an
opaque photosphere. There is no mask applied to the data or the
map. Tests have been done on a PC with two Quad-Core AMD
Opteron$\rm{^{TM}}$ Processor 2380 and 32 Gigabytes of RAM.

In a more realistic use of \textsf{TomograPy} for solar tomography, the
projections would be faster than presented here since masked pixels
are not projected and LOS integration is stopped when the ray
reaches the photosphere, reducing the number of computations.

Performance as a function of the number of threads used by OpenMP are
shown in Table \ref{tab:ncores}. It shows that the time to compute a
projection and a back-projection is not exactly linear with the number
of threads. It takes seven times less time to compute a projection
with eight threads than with one thread and 6.1 times less to compute
a back-projection. Back-projection does not scale as well as
projection (the speed-up with multiple cores is better with the
projection). This is due to the fact that extra care must be taken
while updating the map of voxel values as opposed to the detector
values in order to avoid race conditions. Race conditions are
situation in which the outcome of a computation varies unexpectedly
due to the timing of events in different threads. In
\textsf{TomograPy}, race conditions occur mainly when two LOS need to
update the same voxel at the same time. In the case of the projection,
voxel values are only read, so this not an issue. Race conditions are
avoided with the OpenMP atomic pragma directive which instruct each
thread to update voxels sequentially, resulting in a slow-down of
computations, but only for this part of the algorithm. Note however
that the projections and back-projections scale better when the ratio
between the number of data samples and the number of voxels
decreases. Indeed, in this case, the number of lines of sight
intersecting a single voxel decreases so that it is less probable that
several threads try to update the same voxel at the same time.
\begin{table}
  \small
  \begin{tabular}{p{.28\linewidth}*{8}{c}}
    \hline
    Cores & 1 & 2 & 3 & 4 & 5 & 6 & 7 & 8 \\
    \hline
    Projection time [s] & 97.5 & 48.6 & 36.3 & 26.4 & 21.2 & 18.2 & 16.8 & 13.9 \\
    back-projection time [s] & 145.3 & 76.8 & 56.5 & 41.0 & 37.0 & 32.7 & 30.1 & 23.9 \\
    \hline
  \end{tabular}
  \caption{Performances of the projector as a function of threads.}
  \label{tab:ncores}
\end{table}

Performance as a function of the image format is shown in Table
\ref{tab:image_shape}. As expected, the projection and
back-projection duration scales linearly with the image size.
\begin{table}
  \begin{tabular}{p{.25\linewidth}*{5}{c}}
    \hline
    Image size & 128 $\times$ 128 & 256 $\times$ 256 & 512 $\times$ 512 & 1024 $\times$ 1024 \\
    \hline
    Projection time [s] & 0.95 & 3.64 & 14.53 & 58.75 \\
    back-projection time [s] & 1.61 & 6.17 & 24.69 & 99.54 \\
    \hline
  \end{tabular}
  \caption{Performance of the projector as a function of image shape.}
  \label{tab:image_shape}
\end{table}

Performance as a function of the reconstruction cube size is shown in
Table \ref{tab:cube_shape}. We can see that the projection and
back-projection duration scales linearly with the cube root of the
number of voxels. This is expected as the number of voxels along a
single LOS is roughly proportional to the number of voxels along one
axis of the map and the number of operations is proportional with the
number of intersections between LOS and voxels.
\begin{table}
  \begin{tabular}{p{.25\linewidth}*{5}{c}}
    \hline
    Cube size & 128$^3$ & 256$^3$ & 512$^3$ & 1024$^3$ \\ 
    \hline
    Projection time [s] & 13.84 & 44.56 & 195.69 & 587.56 \\
    back-projection time [s] & 24.71 & 56.64 & 214.55 & 610.75 \\
    \hline
  \end{tabular}
  \caption{Performance of the projector as a function of the size of the reconstruction cube.}
  \label{tab:cube_shape}
\end{table}


\section{Examples}
\label{sec:exemples}

\subsection{Static Reconstruction using STEREO/EUVI A and B}
We performed a conjugate gradient inversion using data from both EUVI
A and B. To avoid issues due to differences in filters, we rescaled the
EUVI B data to EUVI A levels by dividing by the empirically deduced
values provided in Table \ref{tab:cross_cal_ab}
\begin{table}
  \begin{tabular}{*{5}{c}}
    \hline
    bandwidth & 171 & 195 & 284 & 304 \\
    \hline
    B/A ratio & 0.90 & 0.97 & 0.95 & 1.05 \\
    \hline
  \end{tabular}
  \caption{Ratio of sensitivity between EUVI A and EUVI B as a
    function of wavelength.}
  \label{tab:cross_cal_ab}
\end{table}

This operation is justified since the passbands of the twin
instruments were designed and were measured to be nearly identical. To
first order, the difference in spectral response between EUVI A and B
is a scaling factor. The image pairs where chosen to be simultaneous
with a small lossy-compression factor.
Because of the photosphere opacity, we need a full rotation (four
weeks) to have data isotropic coverage of all parts of the
reconstruction cube. If we focus only on the poles, this is no longer
an issue and we can use only two weeks of data. We can also use the
diversity of points of view provided by STEREO to reduce this
duration.  At the time of observation, the STEREO spacecraft were
separated by 86$^{\circ}$, reducing the acquisition time needed for a
complete coverage of the corona to three weeks instead of a full solar
rotation. For polar regions, the required acquisition time is further
reduced to 11 days. We chose pairs of images regularly spaced from 1
to 15 December 2008, with four pairs of images per day and per
observatory, resulting in 118 images. The estimated 3D map is a cube
of $256 \times 256 \times 256$ voxels centered on the Sun with a width
of three solar radii along each axis. In order to save computation
time while remaining consistent with the resolution of the
reconstruction cube, the images were binned $2\times 2$. The
Carrington rotation rate is assumed.

\begin{figure}
  \center
  \includegraphics[width=.9\linewidth]{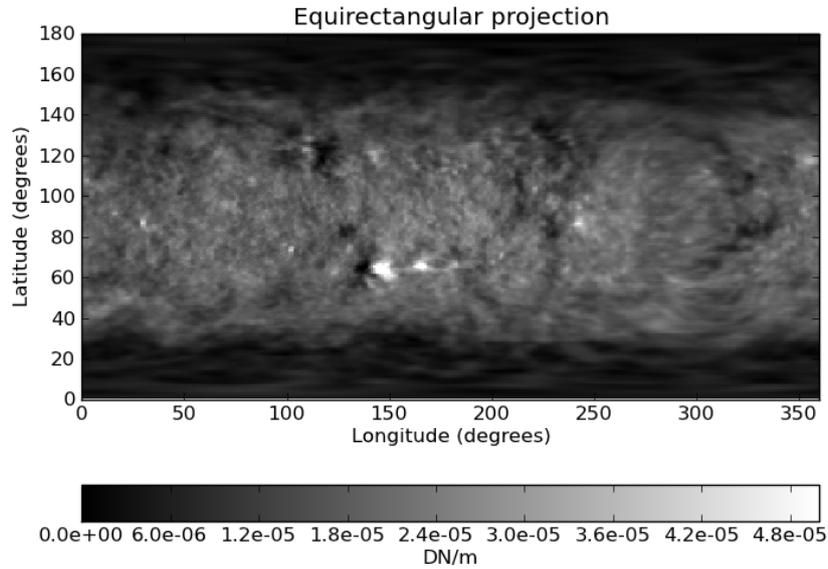}\\
  \caption{Reconstruction of the 17.1 nm local emissivity at 1.05
    solar radii using STEREO/EUVI A and B data and assuming a static
    corona. Input data are 59 pairs of images equally spaced in time
    from 1 to 15 December 2008. The STEREO separation
    angle was 86$^{\circ}$.}
  \label{fig:srt}
\end{figure}

Figure (\ref{fig:srt}) shows the local emissivity in the reconstructed
cube at a constant altitude of 1.05 solar radii. An equi-rectangular
projection is used.

As can be seen, part of the projection around 300\degree{} of
longitude is smoother than the remainder of the projection. This is
due to the fact that we used two weeks of data instead of the three
weeks required to sample all voxels isotropically with lines of
sight. To obtain a better estimate at these locations would have
required the use of one more week's data, which would have in turned
worsened artifacts due to temporal evolution. Conversely, the polar
regions are slightly oversampled because, considering the separation
of the STEREO spacecraft, the minimum integration time was 11
days. Figure \ref{fig:srt_gnomo} shows a gnomonic projection of the
north polar region at 1.05 solar radii. The coronal hole is clearly
visible as the darker central area. Structures in the hole are
arranged according to a network pattern. One can identify bright
nodules that could be attributed to the classical beam plumes but also
elongated structures that could correspond to curtain plumes. This
would confirm the proposition by \inlinecite{gabriel2009} that both
types of plumes coexist.

As can be seen, some of the voxels have negative values. This is
usually explained as resulting from temporal evolution. Indeed, if
temporal evolution has occurred during the acquisition of the data, it
cannot be correctly modeled with the static assumption made in this
inversion. Negative values are thus required to account for a
variation of intensity in the data unexplainable by a simple change of
viewpoint. This could also be explained by mismodeling of the
measurement process, noise or even bias in the data (which could be
due to some instrumental artifcat).

\begin{figure}
  \centering
  \begin{tikzpicture}[x=0.1cm,y=.5cm]

  \def\xmin{0}
  \def\xmax{100}
  \def\ymin{-4}
  \def\ymax{0}
  \draw[style=help lines, ystep=1, xstep=10] (\xmin,\ymin) grid
  (\xmax,\ymax);
  \draw[-] (\xmin,\ymin) -- (\xmax,\ymin) ;
  \draw[-] (\xmin,\ymin) -- (\xmin,\ymax) ;
  \node at (50 , -5) [below] {iteration index};
  \node at (-8., -.5) [left, rotate=90]  {$\log J(\xb)$};
  \foreach \x in {10,20,...,100}
    \node at (\x, \ymin) [below] {\x};
  \foreach \y in {-4,-3,...,0}
    \node at (\xmin,\y) [left] {\y};
  \draw[color=black] plot[mark=*,mark size=1pt] file {critere.txt};

\end{tikzpicture}
  \caption{Example of convergence criterion as a function of the
    iteration index.}
  \label{fig:criterion_exemple}
\end{figure}
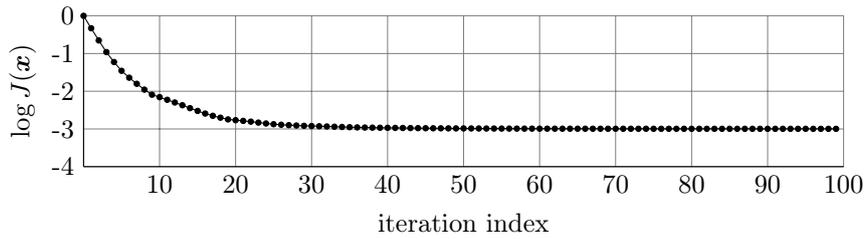

\begin{figure}
  \centering
  \includegraphics[width=.45\linewidth]{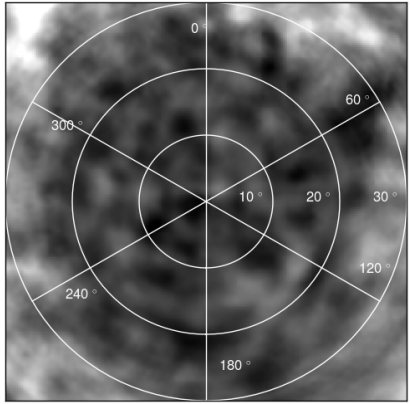}
  \caption{Gonomic projection of the north pole at 1.05 solar radii. The reconstruction cube is the same as that of Figure \ref{fig:srt}. This image can be compared with the dynamic reconstructions of Figure \ref{fig:stsrt}.}
  \label{fig:srt_gnomo}
\end{figure}

To show that convergence has indeed been reached with the conjugate
gradient algorithm we present the criterion as a function of iteration
number in Figure \ref{fig:criterion_exemple}.

\subsection{Smooth Temporal Evolution with STEREO/EUVI A and B}

The second example is a 3D map estimation using the smooth temporal
evolution model. In other words, we minimize Equation (\ref{eq:stsrt})
using the hyper-parameters stated in Table \ref{tab:hypers}. We used
the same set of EUVI images used for the static estimation. For each
pair of EUVI images there is a corresponding instantaneous map of
dimensions $128 \times 128 \times 128$, resulting in the estimation of
approximately 132 millions of parameters. The estimation took less
than eight hours.

\begin{figure}
  \centering
  \includegraphics[width=.9\linewidth]{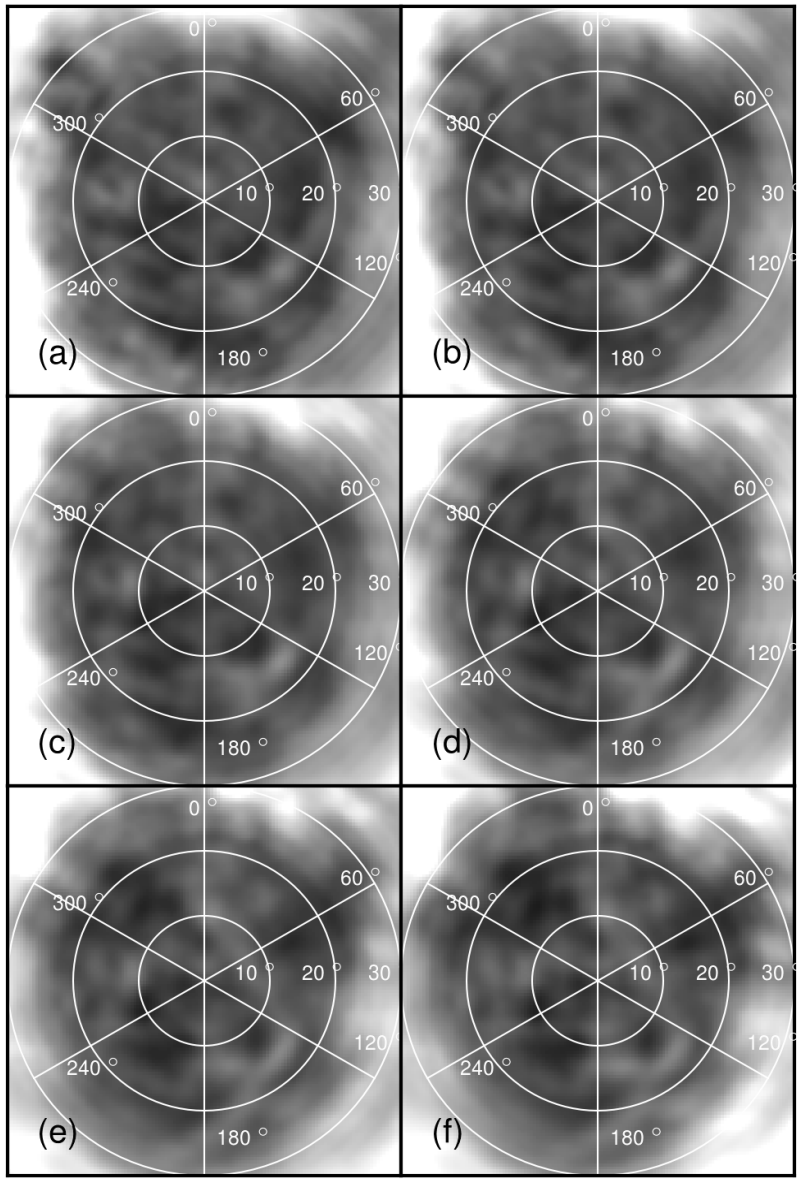}
  \caption{Polar view of the emissivity at 1.05 solar radii from a
    temporally evolving 3D map estimation using STEREO/EUVI A and B
    data in the 17.1 nm passband. North pole gnomonic projection with
    a resolution of 1\degree{} in square is used. There is 40 hours
    between each projection. These images can be compared with the
    static reconstruction of Figure \ref{fig:srt_gnomo}.}
  \label{fig:stsrt}
\end{figure}

Figure \ref{fig:stsrt} shows gnomonic projections at 1.05 solar radii
of the estimated map at different instants separated by 40 hours. It
is interesting to compare these images with the static reconstruction
of the same area shown in Figure \ref{fig:srt_gnomo}. Here, the
disappearance of an elongated structure at the south edge of the
coronal hole is very clear, and one can also follow the appearance and
disappearance of beam plumes.

\begin{table}
  \begin{tabular}{*{5}{c}}
    \hline
    model &$\lambda_x$ & $\lambda_y$ & $\lambda_z$ & $\lambda_t$ \\
    \hline
    SRT & $1e^{-1}$ & $1e^{-1}$ & $1e^{-1}$ & \\
    Thomson & $1e^{-1}$ & $1e^{-1}$ & $1e^{-1}$ & \\
    STSRT & $1e^{-1}$ & $1e^{-1}$ & $1e^{-1}$ & $1e^2$ \\
    \hline
  \end{tabular}
  \caption{Hyper-parameters used for the inversion of the different models.}
  \label{tab:hypers}
\end{table}

The static reconstruction is sharper than the temporally evolving
one. This could be due to slower convergence in the smooth
temporal-rotational tomography model. Indeed, since the temporal prior
is much higher than the spatial prior, numerous very small steps in
the conjugate-gradient algorithm could be required to reach the
minimum. This could be missed by our stopping criterion on the norm of
the gradient and even on convergence diagonstic such as the one in
Figure \ref{fig:criterion_exemple}. This could be solved through the
use of preconditioning but has not been tried for now. However, one
can clearly identify the sames structures in both reconstructions. One
can picture the static reconstruction as a kind of average over time,
although this is not strictly true as temporal effects and changes of
viewpoints can have the same kind of effects on data.

\subsection{Thomson scattering with COR1 A and B data}
\begin{figure}
  \centering
  \includegraphics[width=.9\linewidth]{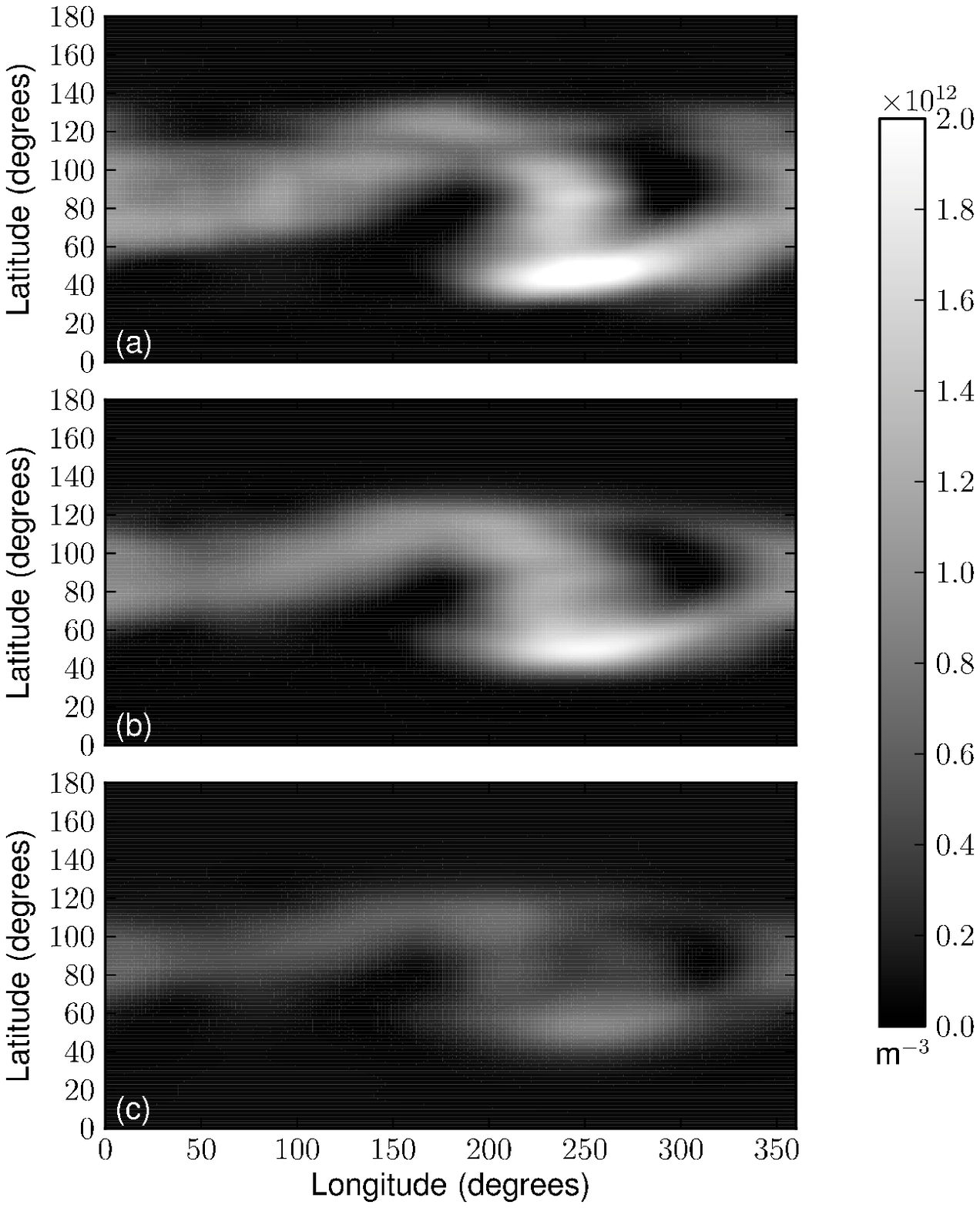}
  \caption{Equi-rectangular projection of a static 3D map estimation
    using pB COR1 A and B data during February 2008 at 1.6 $R_\odot$
    (a), 2.0 $R_\odot$ (b), and 2.4 $R_\odot$ (c). Pixels of the
    projections are 1\degree{} square.  }
  \label{fig:thomson_inv}
\end{figure}

We estimated the coronal electron density using COR1 A and B data
acquired during February 2008 as done by
\inlinecite{kramar2009tomographic}. Since the inversion codes are
different, the comparison gives an estimate of the robustness of both
techniques. Our results are shown in Figure \ref{fig:thomson_inv} and
can be compared to Figure 2 of \inlinecite{kramar2009tomographic}. The
comparison shows that very large scale structures are very similar in
both maps, but fainter and smaller scales structures differ. This can
be explained by the use of different prior models and
hyper-parameters. Data can also differ in the way that they are
prepared before the tomographic inversion. Note that we used a
smoothness prior increasing with height for this reconstruction.

\section{Conclusions}
\label{sec:conclusions}

We developed a tested, fast, and flexible program to perform
rotational tomography of the solar corona. Its respect of WCS
standards allows its use with virtually any data set available.

We demonstrated how this software can be used to perform
multi-spacecraft estimations of emission maps in the corona using
STEREO/EUVI data and STEREO/COR1 data. Estimations were performed
using static, temporally-evolving, and Thomson-scattering models.

This new software suite can naturally be used to perform tomographic
inversion of the corona using SDO/AIA images. \textsf{TomograPy} will
allow tomographic estimation using AIA data at full resolution,
providing unprecedented resolution as well as additional spectral
information.  Supplemented with spectral-inversion methods, this will
allow for electron-density and temperature estimates in large data
sets, close to the photosphere, at very high resolution.

Because of its modular architecture, \textsf{TomograPy} also provides
a convenient way to test different estimation algorithms saving the
need to rewrite anything other than the algorithm. This allows easy
comparison of the performances of different algorithms for the solar
tomography application.

\bibliographystyle{spr-mp-sola}
\bibliography{siddon}

\begin{thebibliography}{36}
\ifx \bisbn   \undefined \def \bisbn  #1{ISBN #1}\fi
\ifx \binits  \undefined \def \binits#1{#1}\fi
\ifx \bauthor  \undefined \def \bauthor#1{#1}\fi
\ifx \batitle  \undefined \def \batitle#1{#1}\fi
\ifx \bjtitle  \undefined \def \bjtitle#1{\textit{#1}}\fi
\ifx \bvolume  \undefined \def \bvolume#1{\textbf{#1}}\fi
\ifx \byear  \undefined \def \byear#1{#1}\fi
\ifx \bissue  \undefined \def \bissue#1{#1}\fi
\ifx \bfpage  \undefined \def \bfpage#1{#1}\fi
\ifx \blpage  \undefined \def \blpage #1{#1}\fi
\ifx \burl  \undefined \def \burl#1{\textsf{#1}}\fi
\ifx \href  \undefined \def \href#1#2{\textsf{#2}}\fi
\ifx \doiurl  \undefined \def
  \doiurl#1{\href{http://dx.doi.org/#1}{\textsf{#1}}}\fi
\ifx \betal  \undefined \def \betal{\textit{et al.}}\fi
\ifx \binstitute  \undefined \def \binstitute#1{#1}\fi
\ifx \bctitle  \undefined \def \bctitle#1{#1}\fi
\ifx \beditor  \undefined \def \beditor#1{#1}\fi
\ifx \bpublisher  \undefined \def \bpublisher#1{#1}\fi
\ifx \bbtitle  \undefined \def \bbtitle#1{\textit{#1}}\fi
\ifx \bedition  \undefined \def \bedition#1{#1}\fi
\ifx \bseriesno  \undefined \def \bseriesno#1{\textbf{#1}}\fi
\ifx \blocation  \undefined \def \blocation#1{#1}\fi
\ifx \bsertitle  \undefined \def \bsertitle#1{\textit{#1}}\fi
\ifx \bsnm \undefined \def \bsnm#1{#1}\fi
\ifx \bsuffix \undefined \def \bsuffix#1{#1}\fi
\ifx \bparticle \undefined \def \bparticle#1{#1}\fi
\ifx \barticle \undefined \def \barticle#1{}\fi
\ifx \botherref \undefined \def \botherref#1{}\fi
\ifx \url \undefined \def \url#1{\textsf{#1}}\fi
\ifx \bchapter \undefined \def \bchapter#1{}\fi
\ifx \bbook \undefined \def \bbook#1{}\fi
\ifx \bcomment \undefined \def \bcomment#1{#1}\fi
\ifx \oauthor \undefined \def \oauthor#1{#1}\fi
\ifx \citeauthoryear \undefined \def \citeauthoryear#1{#1}\fi
\def \endbibitem {}

\bibitem[\protect\citeauthoryear{Altschuler and
  Perry}{1972}]{altschuler1972determining}
\begin{barticle}
\bauthor{\bsnm{Altschuler}, \binits{M.D.}},
\bauthor{\bsnm{Perry}, \binits{R.M.}}:
\byear{1972},
\batitle{{On determining the electron density distribution of the solar corona
  from K-coronameter data}}.
\bjtitle{\solphys}
\bvolume{23}(\bissue{2}),
\bfpage{410}\,--\,\blpage{428}.
\end{barticle}
\endbibitem

\bibitem[\protect\citeauthoryear{Aschwanden
  \textit{et~al.}}{2008}]{aschwanden2008}
\begin{barticle}
\bauthor{\bsnm{Aschwanden}, \binits{M.J.}},
\bauthor{\bsnm{Nitta}, \binits{N.V.}},
\bauthor{\bsnm{Wuelser}, \binits{J.-P.}},
\bauthor{\bsnm{Lemen}, \binits{J.R.}}:
\byear{2008},
\batitle{{First 3D Reconstructions of Coronal Loops with the STEREO A+B
  Spacecraft. II. Electron Density and Temperature Measurements}}.
\bjtitle{\apj}
\bvolume{680}(\bissue{2}),
\bfpage{1477}.
\end{barticle}
\endbibitem

\bibitem[\protect\citeauthoryear{Barbey \textit{et~al.}}{2008}]{barbey2008time}
\begin{barticle}
\bauthor{\bsnm{Barbey}, \binits{N.}},
\bauthor{\bsnm{Auch{\`e}re}, \binits{F.}},
\bauthor{\bsnm{Rodet}, \binits{T.}},
\bauthor{\bsnm{Vial}, \binits{J.C.}}:
\byear{2008},
\batitle{{A Time-Evolving 3D Method Dedicated to the Reconstruction of Solar
  Plumes and Results Using Extreme Ultraviolet Data}}.
\bjtitle{\solphys}
\bvolume{248}(\bissue{2}),
\bfpage{409}.
\end{barticle}
\endbibitem

\bibitem[\protect\citeauthoryear{Barrett and
  Bridgman}{1999}]{barrett1999pyfits}
\bauthor{\bsnm{Barrett}, \binits{P.}},
\bauthor{\bsnm{Bridgman}, \binits{W.}}:
\byear{1999},
\bctitle{{PyFITS, a FITS module for Python}}.
In: \beditor{\bsnm{{D.~M.~Mehringer, R.~L.~Plante, \& D.~A.~Roberts}}} (ed.)
\bbtitle{Astronomical Data Analysis Software and Systems VIII},
\bsertitle{Astronomical Society of the Pacific Conference Series}
\bseriesno{172},
\bfpage{483}.
\endbibitem

\bibitem[\protect\citeauthoryear{{Billings}}{1966}]{billings66}
\begin{bbook}
\bauthor{\bsnm{{Billings}}, \binits{D.E.}}:
\byear{1966},
\bbtitle{{A guide to the solar corona}},
\bpublisher{Academic Press},
\blocation{New York}.
\end{bbook}
\endbibitem

\bibitem[\protect\citeauthoryear{{Brueckner}
  \textit{et~al.}}{1995}]{brueckner1995lasco}
\begin{barticle}
\bauthor{\bsnm{{Brueckner}}, \binits{G.E.}},
\bauthor{\bsnm{{Howard}}, \binits{R.A.}},
\bauthor{\bsnm{{Koomen}}, \binits{M.J.}},
\bauthor{\bsnm{{Korendyke}}, \binits{C.M.}},
\bauthor{\bsnm{{Michels}}, \binits{D.J.}},
\bauthor{\bsnm{{Moses}}, \binits{J.D.}},
\bauthor{\bsnm{{Socker}}, \binits{D.G.}},
\bauthor{\bsnm{{Dere}}, \binits{K.P.}},
\bauthor{\bsnm{{Lamy}}, \binits{P.L.}},
\bauthor{\bsnm{{Llebaria}}, \binits{A.}},
\bauthor{\bsnm{{Bout}}, \binits{M.V.}},
\bauthor{\bsnm{{Schwenn}}, \binits{R.}},
\bauthor{\bsnm{{Simnett}}, \binits{G.M.}},
\bauthor{\bsnm{{Bedford}}, \binits{D.K.}},
\bauthor{\bsnm{{Eyles}}, \binits{C.J.}}:
\byear{1995},
\batitle{{The Large Angle Spectroscopic Coronagraph (LASCO)}}.
\bjtitle{\solphys}
\bvolume{162},
\bfpage{357}.
\end{barticle}
\endbibitem

\bibitem[\protect\citeauthoryear{Butala
  \textit{et~al.}}{2010}]{butala2010dynamic}
\begin{barticle}
\bauthor{\bsnm{Butala}, \binits{M.}},
\bauthor{\bsnm{Hewett}, \binits{R.}},
\bauthor{\bsnm{Frazin}, \binits{R.}},
\bauthor{\bsnm{Kamalabadi}, \binits{F.}}:
\byear{2010},
\batitle{{Dynamic Three-Dimensional Tomography of the Solar Corona}}.
\bjtitle{\solphys}
\bvolume{262}(\bissue{2}),
\bfpage{495}.
\end{barticle}
\endbibitem

\bibitem[\protect\citeauthoryear{Calabretta and
  Greisen}{2002}]{calabretta2002representations}
\begin{barticle}
\bauthor{\bsnm{Calabretta}, \binits{M.R.}},
\bauthor{\bsnm{Greisen}, \binits{E.W.}}:
\byear{2002},
\batitle{{Representations of celestial coordinates in FITS}}.
\bjtitle{\aap}
\bvolume{395},
\bfpage{1077}.
\end{barticle}
\endbibitem

\bibitem[\protect\citeauthoryear{Dagum and Menon}{2002}]{dagum2002openmp}
\begin{barticle}
\bauthor{\bsnm{Dagum}, \binits{L.}},
\bauthor{\bsnm{Menon}, \binits{R.}}:
\byear{2002},
\batitle{{OpenMP: an industry standard API for shared-memory programming}}.
\bjtitle{Comput. Science Eng., IEEE}
\bvolume{5}(\bissue{1}),
\bfpage{46}.
\end{barticle}
\endbibitem

\bibitem[\protect\citeauthoryear{Davila}{1994}]{davila1994solar}
\begin{barticle}
\bauthor{\bsnm{Davila}, \binits{J.M.}}:
\byear{1994},
\batitle{{Solar tomography}}.
\bjtitle{\apj}
\bvolume{423},
\bfpage{871}.
\end{barticle}
\endbibitem

\bibitem[\protect\citeauthoryear{Delaboudini\`ere
  \textit{et~al.}}{1995}]{delaboudiniere1995eit}
\begin{barticle}
\bauthor{\bsnm{Delaboudini\`ere}, \binits{J.P.}},
\bauthor{\bsnm{Artzner}, \binits{G.}},
\bauthor{\bsnm{Brunaud}, \binits{J.}},
\bauthor{\bsnm{Gabriel}, \binits{A.H.}},
\bauthor{\bsnm{Hochedez}, \binits{J.F.}},
\bauthor{\bsnm{Millier}, \binits{F.}},
\bauthor{\bsnm{Song}, \binits{X.Y.}},
\bauthor{\bsnm{Au}, \binits{B.}},
\bauthor{\bsnm{Dere}, \binits{K.P.}},
\bauthor{\bsnm{Howard}, \binits{R.A.}}, \betal:
\byear{1995},
\batitle{{EIT: extreme-ultraviolet imaging telescope for the SOHO mission}}.
\bjtitle{\solphys}
\bvolume{162}(\bissue{1}),
\bfpage{291}.
\end{barticle}
\endbibitem

\bibitem[\protect\citeauthoryear{Frazin, V\'asquez, and
  Kamalabadi}{2009}]{frazin2009}
\begin{barticle}
\bauthor{\bsnm{Frazin}, \binits{R.A.}},
\bauthor{\bsnm{V\'asquez}, \binits{A.M.}},
\bauthor{\bsnm{Kamalabadi}, \binits{F.}}:
\byear{2009},
\batitle{{Quantitative, Three-dimensional Analysis of the Global Corona with
  Multi-spacecraft Differential Emission Measure Tomography}}.
\bjtitle{\apj}
\bvolume{701},
\bfpage{547}.
\end{barticle}
\endbibitem

\bibitem[\protect\citeauthoryear{Frazin}{2000}]{frazin2000tomography}
\begin{barticle}
\bauthor{\bsnm{Frazin}, \binits{R.A.}}:
\byear{2000},
\batitle{{Tomography of the solar corona. I. A robust, regularized, positive
  estimation method}}.
\bjtitle{\apj}
\bvolume{530},
\bfpage{1026}.
\end{barticle}
\endbibitem

\bibitem[\protect\citeauthoryear{Frazin and
  Janzen}{2002}]{frazin2002tomography}
\begin{barticle}
\bauthor{\bsnm{Frazin}, \binits{R.A.}},
\bauthor{\bsnm{Janzen}, \binits{P.}}:
\byear{2002},
\batitle{{Tomography of the solar corona. II. Robust, regularized, positive
  estimation of the three-dimensional electron density distribution from
  LASCO-C2 polarized white-light images}}.
\bjtitle{\apj}
\bvolume{570},
\bfpage{408}.
\end{barticle}
\endbibitem

\bibitem[\protect\citeauthoryear{Frazin and
  Kamalabadi}{2005}]{frazin2005rotational}
\begin{barticle}
\bauthor{\bsnm{Frazin}, \binits{R.A.}},
\bauthor{\bsnm{Kamalabadi}, \binits{F.}}:
\byear{2005},
\batitle{{Rotational tomography for 3D reconstruction of the white-light and
  EUV corona in the post-SOHO era}}.
\bjtitle{\solphys}
\bvolume{228}(\bissue{1}),
\bfpage{219}.
\end{barticle}
\endbibitem

\bibitem[\protect\citeauthoryear{Frazin
  \textit{et~al.}}{2010}]{frazin2010three}
\begin{botherref}
\oauthor{\bsnm{Frazin}, \binits{R.A.}},
\oauthor{\bsnm{Lamy}, \binits{P.}},
\oauthor{\bsnm{Llebaria}, \binits{A.}},
\oauthor{\bsnm{V{\'a}squez}, \binits{A.M.}}:
2010,
{Three-Dimensional Electron Density from Tomographic Analysis of LASCO-C2
  Images of the K-Corona Total Brightness}.
\textit{Solar Physics},
1\,--\,12.
\end{botherref}
\endbibitem

\bibitem[\protect\citeauthoryear{Gabriel \textit{et~al.}}{2009}]{gabriel2009}
\begin{barticle}
\bauthor{\bsnm{Gabriel}, \binits{A.H.}},
\bauthor{\bsnm{Bely-Dubau}, \binits{F.}},
\bauthor{\bsnm{Tison}, \binits{E.}},
\bauthor{\bsnm{Wilhelm}, \binits{K.}}:
\byear{2009},
\batitle{{The structure and origin of solar plumes: network plumes}}.
\bjtitle{\apj}
\bvolume{700},
\bfpage{551}.
\end{barticle}
\endbibitem

\bibitem[\protect\citeauthoryear{Greisen and
  Calabretta}{2002}]{greisen2002representations}
\begin{barticle}
\bauthor{\bsnm{Greisen}, \binits{E.W.}},
\bauthor{\bsnm{Calabretta}, \binits{M.R.}}:
\byear{2002},
\batitle{{Representations of world coordinates in FITS}}.
\bjtitle{\aap}
\bvolume{395},
\bfpage{1061}.
\end{barticle}
\endbibitem

\bibitem[\protect\citeauthoryear{Jones, Oliphant, and Peterson}{2001--}]{scipy}
\begin{botherref}
\oauthor{\bsnm{Jones}, \binits{E.}},
\oauthor{\bsnm{Oliphant}, \binits{T.}},
\oauthor{\bsnm{Peterson}, \binits{P.}}:
2001--,
\textit{{SciPy}: Open source scientific tools for {Python}}.
\url{http://www.scipy.org/}.
\end{botherref}
\endbibitem

\bibitem[\protect\citeauthoryear{Kaiser
  \textit{et~al.}}{2008}]{kaiser2008stereo}
\begin{barticle}
\bauthor{\bsnm{Kaiser}, \binits{M.}},
\bauthor{\bsnm{Kucera}, \binits{T.}},
\bauthor{\bsnm{Davila}, \binits{J.}},
\bauthor{\bsnm{St.~Cyr}, \binits{O.}},
\bauthor{\bsnm{Guhathakurta}, \binits{M.}},
\bauthor{\bsnm{Christian}, \binits{E.}}:
\byear{2008},
\batitle{{The STEREO mission: An introduction}}.
\bjtitle{\ssr}
\bvolume{136}(\bissue{1}),
\bfpage{5}.
\end{barticle}
\endbibitem

\bibitem[\protect\citeauthoryear{Khalsa and
  Fessler}{2007}]{khalsa2007resolution}
\bauthor{\bsnm{Khalsa}, \binits{K.A.}},
\bauthor{\bsnm{Fessler}, \binits{J.A.}}:
\byear{2007},
\bctitle{{Resolution properties in regularized dynamic MRI reconstruction}}.
In: \bbtitle{Biomedical Imaging: From Nano to Macro, 2007. ISBI 2007. 4th IEEE
  International Symposium on},
\bfpage{456}\,--\,\blpage{459}.
\bcomment{IEEE}.
\bisbn{1424406722}.
\endbibitem

\bibitem[\protect\citeauthoryear{Kohl \textit{et~al.}}{1995}]{kohl1995}
\begin{barticle}
\bauthor{\bsnm{Kohl}, \binits{J.L.}},
\bauthor{\bsnm{Esser}, \binits{R.}},
\bauthor{\bsnm{Gardner}, \binits{L.D.}},
\bauthor{\bsnm{Habbal}, \binits{S.}},
\bauthor{\bsnm{Daigneau}, \binits{P.S.}},
\bauthor{\bsnm{Dennis}, \binits{E.F.}},
\bauthor{\bsnm{Nystrom}, \binits{G.U.}},
\bauthor{\bsnm{Panasyuk}, \binits{A.}},
\bauthor{\bsnm{Raymond}, \binits{J.C.}},
\bauthor{\bsnm{Smith}, \binits{P.L.}},
\bauthor{\bsnm{Strachan}, \binits{L.}},
\bauthor{\bparticle{van} \bsnm{Ballegooijen}, \binits{A.A.}},
\bauthor{\bsnm{Noci}, \binits{G.}},
\bauthor{\bsnm{Fineschi}, \binits{S.}},
\bauthor{\bsnm{Romoli}, \binits{M.}},
\bauthor{\bsnm{Ciaravella}, \binits{A.}},
\bauthor{\bsnm{Modigliani}, \binits{A.}},
\bauthor{\bsnm{Huber}, \binits{M.C.E.}},
\bauthor{\bsnm{Antonucci}, \binits{E.}},
\bauthor{\bsnm{Benna}, \binits{C.}},
\bauthor{\bsnm{Giordano}, \binits{S.}},
\bauthor{\bsnm{Tondello}, \binits{G.}},
\bauthor{\bsnm{Nicolosi}, \binits{P.}},
\bauthor{\bsnm{Naletto}, \binits{G.}},
\bauthor{\bsnm{Pernechele}, \binits{C.}},
\bauthor{\bsnm{Spadaro}, \binits{D.}},
\bauthor{\bsnm{Poletto}, \binits{G.}},
\bauthor{\bsnm{Livi}, \binits{S.}},
\bauthor{\bparticle{von~der} \bsnm{L\"uhe}, \binits{O.}},
\bauthor{\bsnm{Geiss}, \binits{J.}},
\bauthor{\bsnm{Timothy}, \binits{J.G.}},
\bauthor{\bsnm{Gloeckler}, \binits{G.}},
\bauthor{\bsnm{Allegra}, \binits{A.}},
\bauthor{\bsnm{Basile}, \binits{G.}},
\bauthor{\bsnm{Brusa}, \binits{R.}},
\bauthor{\bsnm{Wood}, \binits{B.}},
\bauthor{\bsnm{Siegmund}, \binits{O.H.W.}},
\bauthor{\bsnm{Fowler}, \binits{W.}},
\bauthor{\bsnm{Fisher}, \binits{R.}},
\bauthor{\bsnm{Jhabvala}, \binits{M.}}:
\byear{1995},
\batitle{{The Ultraviolet Coronagraph Spectrometer for the Solar and
  Heliospheric Observatory}}.
\bjtitle{\solphys}
\bvolume{162},
\bfpage{313}.
\end{barticle}
\endbibitem

\bibitem[\protect\citeauthoryear{Kramar
  \textit{et~al.}}{2009}]{kramar2009tomographic}
\begin{barticle}
\bauthor{\bsnm{Kramar}, \binits{M.}},
\bauthor{\bsnm{Jones}, \binits{S.}},
\bauthor{\bsnm{Davila}, \binits{J.}},
\bauthor{\bsnm{Inhester}, \binits{B.}},
\bauthor{\bsnm{Mierla}, \binits{M.}}:
\byear{2009},
\batitle{{On the Tomographic Reconstruction of the 3D Electron Density for the
  Solar Corona from STEREO COR1 Data}}.
\bjtitle{\solphys}
\bvolume{259}(\bissue{1}),
\bfpage{109}.
\end{barticle}
\endbibitem

\bibitem[\protect\citeauthoryear{Oliphant}{2006}]{oliphant2006guide}
\begin{bbook}
\bauthor{\bsnm{Oliphant}, \binits{T.E.}}:
\byear{2006},
\bbtitle{{A Guide to NumPy}}
\bseriesno{1},
\bpublisher{Trelgol Publishing},
\blocation{USA}.
\end{bbook}
\endbibitem

\bibitem[\protect\citeauthoryear{Panasyuk}{1999}]{panasyuk1999three}
\begin{barticle}
\bauthor{\bsnm{Panasyuk}, \binits{A.V.}}:
\byear{1999},
\batitle{{Three-dimensional reconstruction of UV emissivities in the solar
  corona using Ultraviolet Coronagraph Spectrometer data from the Whole Sun
  Month}}.
\bjtitle{\jgr}
\bvolume{104}(\bissue{A5}),
\bfpage{9721}.
\end{barticle}
\endbibitem

\bibitem[\protect\citeauthoryear{Schrijver and McMullen}{2000}]{schrijver2000}
\begin{barticle}
\bauthor{\bsnm{Schrijver}, \binits{C.J.}},
\bauthor{\bsnm{McMullen}, \binits{R.A.}}:
\byear{2000},
\batitle{{A Case for Resonant Scattering in the Quiet Solar Corona in
  Extreme-Ultraviolet Lines with High Oscillator Strengths}}.
\bjtitle{\apj}
\bvolume{531},
\bfpage{1121}.
\end{barticle}
\endbibitem

\bibitem[\protect\citeauthoryear{Siddon}{1985}]{siddon85}
\begin{barticle}
\bauthor{\bsnm{Siddon}, \binits{R.L.}}:
\byear{1985},
\batitle{{Fast calculation of the exact radiological path for a
  three-dimensional {CT} array}}.
\bjtitle{Medical Phys.}
\bvolume{12},
\bfpage{252}\,--\,\blpage{255}.
\end{barticle}
\endbibitem

\bibitem[\protect\citeauthoryear{Soleimani and
  Lionheart}{2005}]{soleimani2005nonlinear}
\begin{barticle}
\bauthor{\bsnm{Soleimani}, \binits{M.}},
\bauthor{\bsnm{Lionheart}, \binits{W.R.B.}}:
\byear{2005},
\batitle{{Nonlinear image reconstruction for electrical capacitance tomography
  using experimental data}}.
\bjtitle{Measurement Science and Technology}
\bvolume{16},
\bfpage{1987}.
\end{barticle}
\endbibitem

\bibitem[\protect\citeauthoryear{Terzo and Reale}{2010}]{terzo2010}
\begin{barticle}
\bauthor{\bsnm{Terzo}, \binits{S.}},
\bauthor{\bsnm{Reale}, \binits{F.}}:
\byear{2010},
\batitle{{On the importance of background subtraction in the analysis of
  coronal loops observed with TRACE}}.
\bjtitle{\aap}
\bvolume{515},
\bfpage{A7}.
\end{barticle}
\endbibitem

\bibitem[\protect\citeauthoryear{Thernisien, Vourlidas, and
  Howard}{2009}]{thernisien2009forward}
\begin{barticle}
\bauthor{\bsnm{Thernisien}, \binits{A.}},
\bauthor{\bsnm{Vourlidas}, \binits{A.}},
\bauthor{\bsnm{Howard}, \binits{R.}}:
\byear{2009},
\batitle{{Forward modeling of coronal mass ejections using STEREO/SECCHI
  data}}.
\bjtitle{Solar Physics}
\bvolume{256}(\bissue{1}),
\bfpage{111}\,--\,\blpage{130}.
\end{barticle}
\endbibitem

\bibitem[\protect\citeauthoryear{Van~de Hulst}{1950}]{van1950electron}
\begin{barticle}
\bauthor{\bparticle{Van~de} \bsnm{Hulst}, \binits{H.}}:
\byear{1950},
\batitle{{The electron density of the solar corona}}.
\bjtitle{Bull. Astronom. Inst. Netherlands}
\bvolume{11},
\bfpage{135}.
\end{barticle}
\endbibitem

\bibitem[\protect\citeauthoryear{Van~Rossum and {Centrum voor Wiskunde en
  Informatica}}{1995}]{van1995python}
\begin{bbook}
\bauthor{\bsnm{Van~Rossum}, \binits{G.}},
\bauthor{\bsnm{{Centrum voor Wiskunde en Informatica}}}:
\byear{1995},
\bbtitle{{Python reference manual}},
\bpublisher{Centrum voor Wiskunde en Informatica},
\blocation{Amsterdam}.
\end{bbook}
\endbibitem

\bibitem[\protect\citeauthoryear{{Wells}, {Greisen}, and
  {Harten}}{1981}]{wells81}
\begin{barticle}
\bauthor{\bsnm{{Wells}}, \binits{D.C.}},
\bauthor{\bsnm{{Greisen}}, \binits{E.W.}},
\bauthor{\bsnm{{Harten}}, \binits{R.H.}}:
\byear{1981},
\batitle{{FITS - a Flexible Image Transport System}}.
\bjtitle{Astron. and Astrophys. Supp. Ser.}
\bvolume{44},
\bfpage{363}.
\end{barticle}
\endbibitem

\bibitem[\protect\citeauthoryear{Wiegelmann and
  Inhester}{2003}]{wiegelmann2003magnetic}
\begin{barticle}
\bauthor{\bsnm{Wiegelmann}, \binits{T.}},
\bauthor{\bsnm{Inhester}, \binits{B.}}:
\byear{2003},
\batitle{{Magnetic modeling and tomography: First steps towards a consistent
  reconstruction of the solar corona}}.
\bjtitle{\solphys}
\bvolume{214}(\bissue{2}),
\bfpage{287}.
\end{barticle}
\endbibitem

\bibitem[\protect\citeauthoryear{Wuelser
  \textit{et~al.}}{2004}]{wuelser2004euvi}
\bauthor{\bsnm{Wuelser}, \binits{J.P.}},
\bauthor{\bsnm{Lemen}, \binits{J.R.}},
\bauthor{\bsnm{Tarbell}, \binits{T.D.}},
\bauthor{\bsnm{Wolfson}, \binits{C.}},
\bauthor{\bsnm{Cannon}, \binits{J.C.}},
\bauthor{\bsnm{Carpenter}, \binits{B.A.}},
\bauthor{\bsnm{Duncan}, \binits{D.W.}},
\bauthor{\bsnm{Gradwohl}, \binits{G.S.}},
\bauthor{\bsnm{Meyer}, \binits{S.B.}},
\bauthor{\bsnm{Moore}, \binits{A.S.}}, \betal:
\byear{2004},
\bctitle{{EUVI: the STEREO-SECCHI extreme ultraviolet imager}}.
In: \bbtitle{Proceedings of SPIE}
\bseriesno{5171},
\bfpage{111}.
\endbibitem

\bibitem[\protect\citeauthoryear{Zhang, Ghodrati, and
  Brooks}{2005}]{zhang2005analytical}
\begin{barticle}
\bauthor{\bsnm{Zhang}, \binits{Y.}},
\bauthor{\bsnm{Ghodrati}, \binits{A.}},
\bauthor{\bsnm{Brooks}, \binits{D.H.}}:
\byear{2005},
\batitle{{An analytical comparison of three spatio-temporal regularization
  methods for dynamic linear inverse problems in a common statistical
  framework}}.
\bjtitle{Inverse Problems}
\bvolume{21},
\bfpage{357}.
\end{barticle}
\endbibitem

\end{thebibliography}
\end{article}
\end{document}